\documentclass[12pt,oneside,notitlepage,a4paper]{article}
\usepackage[polutonikogreek,english]{babel}
\usepackage{amsmath}
\usepackage{amssymb}
\usepackage{amsfonts}
\usepackage{amsthm}
\usepackage{graphicx}
\usepackage[font=small,labelsep=endash]{caption}
\usepackage{hhline}
\usepackage{subfig}
\usepackage{color}
\usepackage{calc}
\usepackage{enumerate}
\usepackage[round]{natbib}

\def\frac#1#2{{\textstyle{#1\over#2}}}

\def\sign{{\rm sign}}

\DeclareSymbolFont{AMSb}{U}{msb}{m}{n}
\DeclareMathSymbol{\Natural}{\mathbin}{AMSb}{"4E}
\DeclareMathSymbol{\Integer}{\mathbin}{AMSb}{"5A}
\DeclareMathSymbol{\Real}{\mathbin}{AMSb}{"52}
\DeclareMathSymbol{\Rational}{\mathbin}{AMSb}{"51}
\DeclareMathSymbol{\Imaginary}{\mathbin}{AMSb}{"49}
\DeclareMathSymbol{\Complex}{\mathbin}{AMSb}{"43} 
\def\bi{\begin{itemize}}
\def\ei{\end{itemize}}
\def\bd{\begin{description}}
\def\ed{\end{description}}
\def\ben{\begin{enumerate}}
\def\een{\end{enumerate}}

\def\d{{\delta}}

\def\bar#1{{\overline{#1}}}
\def\hat#1{{\widehat{#1}}}

\def\pr{{\rm Pr}}
\def\Pr{\pr}
\def\E{{\rm E}}

\def\var{{\rm var}}

\def\2to{{\ {\buildrel 2\over \longrightarrow}\ }}

\def\I1ton{{$I_1,\ldots,I_n$}}
\def\X1ton{{$X_1,\ldots,X_n$}}
\def\Y1ton{{$Y_1,\ldots,Y_n$}}
\def\Z1ton{{$Z_1,\ldots,Z_n$}}
\def\R1ton{{$R_1,\ldots,R_n$}}
\def\e1ton{{$e_1,\ldots,e_n$}}
\def\t1ton{{$t_1,\ldots,t_n$}}
\def\x1ton{{$x_1,\ldots,x_n$}}
\def\y1ton{{$y_1,\ldots,y_n$}}
\def\z1ton{{$z_1,\ldots,z_n$}}
\def\r1ton{{$r_1,\ldots,r_n$}}

\def\tt#1{{\texttt{#1}}}

\usepackage[pdftex, colorlinks,%
linkcolor=black, citecolor=black,%
unicode, bookmarks=false, hypertexnames=false]{hyperref}

\renewcommand{\textwidth}{6.3in}
\renewcommand{\textheight}{21cm}
\newcommand{\gras}{\boldsymbol}

\numberwithin{equation}{section}

\renewcommand{\d}[1]{\ensuremath{\operatorname{d}\!{#1}}}

\begin{document}
\title{\vspace*{-11ex} Bayesian Uncertainty Management in Temporal Dependence of Extremes}
\author{T.~Lugrin, A.~C.~Davison and J.~A.~Tawn%
\footnote{Thomas Lugrin is a PhD candidate and Anthony Davison is Professor of Statistics, EPFL-FSB-MATHAA-STAT, Station 8, Ecole Polytechnique F\'ed\'erale de Lausanne, 1015 Lausanne, Switzerland, {\tt Thomas.Lugrin@epfl.ch}, {\tt Anthony.Davison@epfl.ch}.  Jonathan Tawn is Professor of Statistics, Department of Mathematics and Statistics, Lancaster University, Lancaster LA1 4YF, UK, \texttt{j.tawn@lancs.ac.uk}.
This research was partially supported by the Swiss National Science Foundation.
}
}
\date{\today}
\maketitle

\begin{abstract}
Both marginal and dependence features must be described when modelling the extremes of a stationary time series. There are standard approaches to marginal modelling, but long- and short-range dependence of extremes may both appear. In applications, an assumption of long-range independence often seems reasonable, but short-range dependence, i.e., the clustering of extremes, needs attention.  The extremal index $0<\theta\le 1$ is a natural limiting measure of clustering, but for wide classes of dependent processes, including all stationary Gaussian processes, it cannot distinguish dependent processes from independent processes with $\theta=1$. \citet{EastoeTawn} exploit methods from multivariate extremes to treat the subasymptotic extremal dependence structure of stationary time series, covering both $0<\theta<1$ and $\theta=1$, through the introduction of a threshold-based extremal index.  Inference for their dependence models uses an inefficient stepwise procedure that has various weaknesses and has no reliable assessment of uncertainty. We overcome these issues using a Bayesian semiparametric approach.  Simulations and the analysis of a UK daily river flow time series show that the new approach provides improved efficiency for estimating properties of functionals of clusters.

\end{abstract}
\newpage


\section{Introduction}\label{sec:intro}


Extreme value theory provides an asymptotically justified framework for the statistical modelling of rare events. In the univariate case with independent variables there is a broadly-used framework involving modelling exceedances of a high threshold by a generalised Pareto distribution \citep{ismev2001}. For extremes of stationary univariate time series, standard procedures use marginal extreme value modelling but consideration of the dependence structure between the variables is essential when assessing risk due to clusters of extremes.
\citet{LeadbetterLindgrenRootzen83} and \citet{HsingHuslerLeadbetter88} described the roles of long- and short-range dependence on extremes of stationary time series. Typically an assumption of independence at long range is reasonable, with \citet{Ledford03} giving diagnostic methods for testing this. In practice independent clusters are often identified using  the runs method \citep{SmithWeissman}, which deems successive exceedances to be in separate clusters  if they are separated by at least $m$ consecutive non-exceedances of the threshold;  \citet{FerroSegers} provide automatic methods for the selection of $m$. 

Short-range dependence has the most important practical implications, since it leads to local temporal clustering of extreme values. \citet{Leadbetter83} and \citet{OBrien} provide different asymptotic characterisations of the clustering though the extremal index $0<\theta\le 1$, the former with $\theta^{-1}$ being the limiting mean cluster size. The case $\theta=1$ corresponds to there being no more clustering than if the series were independent, and decreasing $\theta$ corresponds to increased clustering.

Although the extremal index is a natural limiting measure for such clustering there is a broad class of dependent processes with $\theta=1$, including all stationary Gaussian processes. Thus the extremal index cannot distinguish the clustering properties of this class of dependent processes from those of white noise. Furthermore, many other functionals of clusters of extremes may be of interest \citep{Segers03}, so for practical modelling  of the clusters of extreme values above a high threshold, the restriction to $\theta<1$ is a major weakness.

\cite{SmithTawnColes97}, \cite{Ledford03}, \cite{EastoeTawn} and \citet{WinterTawn} draw on methodology for multivariate extremes to provide models for univariate clustering, and thereby enable the properties of a wide range of cluster functionals to be estimated.  The focus of this paper is the improvement of inference techniques for the most general model for these cases, namely the semiparametric conditional extremes model of \citet{HeffernanTawn}. Our ideas can be applied to any cluster functional, but we focus here primarily on the threshold-based extremal index introduced by \citet{Ledford03}.

Given the strong connections between multivariate extreme value and clustering modelling, here and in Section~\ref{sec:SUB} we present the developments of the model in parallel for the two situations.  Examples of applications for multivariate cases include assessing the risk of joint occurrence of extreme river flows or sea-levels at different locations \citep{Keef1,AsadiDavisonEngelke}, the concurrent high levels of different pollutants at the same location \citep{HeffernanTawn}, and simultaneous stock market crashes \citep{PoonRockingerTawn}. For the time series case, applications include assessing heatwave risks \citep{ReichShabyCooley,WinterTawn}, modelling of extreme rainfall events \citep{SuvegesDavison} and wind gusts \citep{FawcettWalshaw06}.

\begin{figure}[!t]
\centering
\includegraphics[width=12cm, clip=true, trim=0cm 0cm 0cm 1cm]{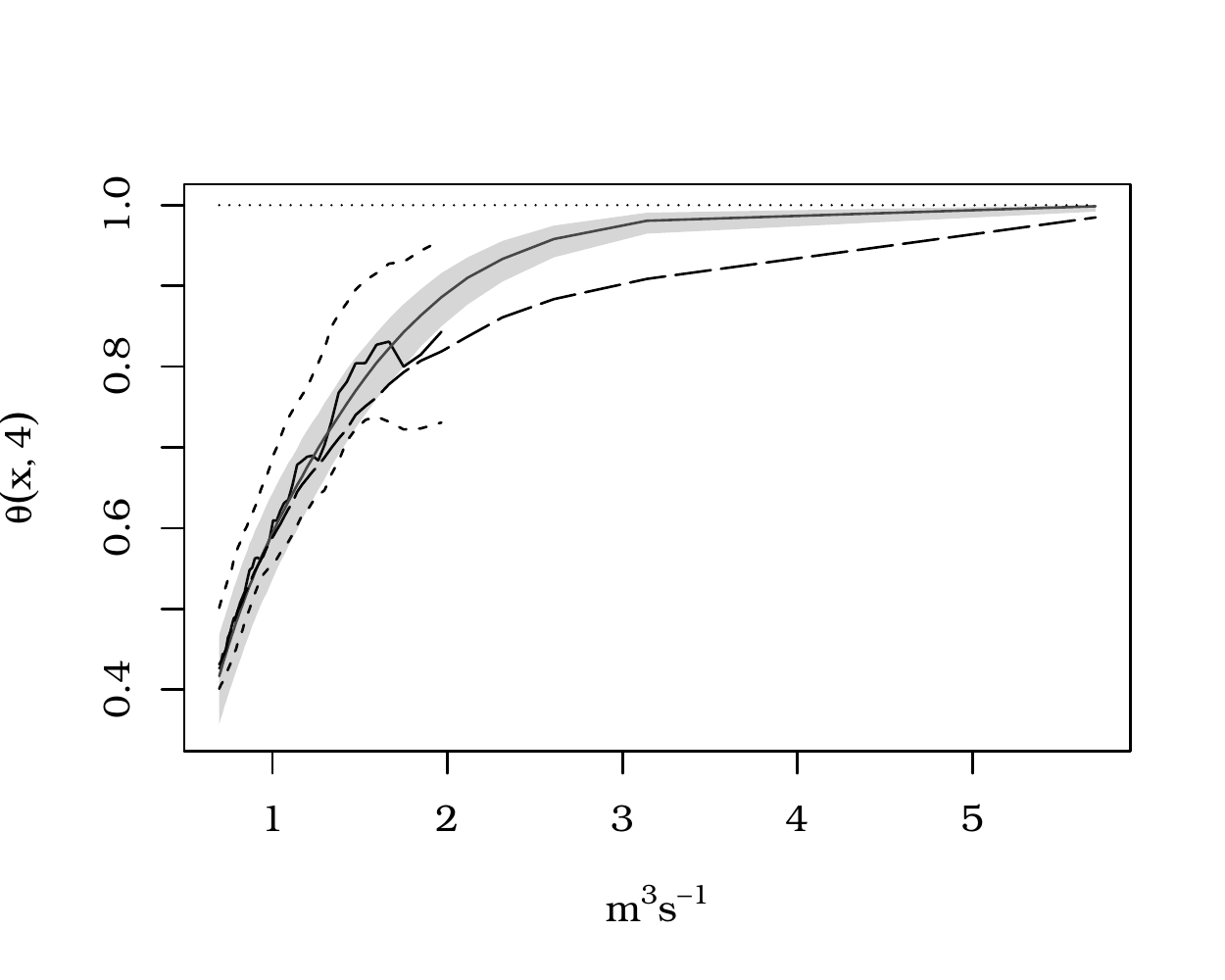}
\caption{Comparison of the empirical, stepwise and Bayesian semiparametric estimates of $\theta(x,4)$ described in Sections~\ref{sec:SUB} and \ref{sec:BAY} respectively. Empirical estimate (solid black) and stepwise (long dashed) and Bayesian (solid grey) semiparametric estimates with their respective $95\%$ confidence regions (dashed and shaded), estimated on winter flows (m$^3$s$^{-1}$) of the River Ray from $1962$ to $2008$.}
\label{fig:theta}
\end{figure}

For the stationary time series $(X_t)$, \citet{Ledford03} define the threshold-based extremal index
\begin{equation}
\label{theta-x-m.eq}
\theta(x,m) = \pr\left(X_1\leq x,\ldots,X_m\leq x\mid X_0>x\right),
\end{equation}
where $x$ is large, which is the key measure of short-range clustering of extreme values, with $1/\theta(x,m)$ being the mean cluster size when clusters of exceedances of the threshold $x$ are defined via the runs method of \citet{SmithWeissman} with run length $m$. Furthermore $\theta(x,m)$ converges to the extremal index $\theta$ as $x\to x_F$ and $m\to\infty$ appropriately \citep{OBrien,KratzRootzen97}.
Many studies have focused on estimating the limit $\theta$ \citep{FerroSegers,Suveges,CYRobert}, but in applications, all these consider a finite level $u$ as an approximation to the limit $x_F$. This is equivalent to assuming $\theta(x,m)$ to be constant above $u$, which is generally not the case in applications (see Figure~\ref{fig:theta}).
Additionally \citet{EastoeTawn} find that $\theta(x,m)$ is fundamental to modelling the distributions of both cluster maxima, i.e., peaks over threshold, and block maxima, e.g., annual maxima. See Section~\ref{sec:ts} for more on the relevance of $\theta(x,m)$ for time series extremes.

When considering asymptotically motivated models for the joint distribution of ${\bf X}_{-0}=(X_1,\ldots,X_m)$ given that $X_0>x$ for the estimation of $\theta(x,m)$, it is helpful to have a simple characterisation of extremal dependence. The standard pairwise measure of extremal dependence for $(X_0,X_j)$ is
\begin{equation}\label{eq:chi}
\chi_j=\lim_{x\to x_F}\pr\left(X_j>x\mid X_0>x\right),\quad j=1,\ldots, m, 
\end{equation}
with the cases $\chi_j>0$ and $\chi_j=0$  respectively termed asymptotic dependence and asymptotic independence at lag $j$. A plot of $\chi_j$ against $j$ has been termed the extremogram \citep{DavisMikosh}, by analogy with the correlogram of time series analysis.  When $\chi_j=0$ for all $j\ge 1$, the extremogram fails to distinguish between different levels of asymptotic independence, but the rate of convergence to zero of $\pr(X_j>x\mid X_0>x)$ determines the key characteristics of the tail of the joint distribution \citep{Ledford96}. \citet{Ledford03} propose using such a measure at each time lag $j$ when the variables are asymptotically independent. An alternative is to combine both approaches by studying a threshold-based version of $\chi_j$, 
\begin{equation}\label{eq:chix}
\chi_j(x)=\pr\left(X_j>x\mid X_0>x\right),\quad j=1,\ldots, m, 
\end{equation}
for a range of large values of $x$.


In Section~\ref{sec:history}, we review classical multivariate extreme models, which all entail $\chi_j(x)=\chi_j$, $x>u$ for some high threshold $u$, and often even $\chi_j>0$. Instead we consider the conditional formulation of \citet{HeffernanTawn} that has been subsequently studied more theoretically by \citet{HeffernanResnick}, \citet{DasResnick}, and \citet{MitraResnick}. This class of models covers $\chi_j\geq 0$ and $\chi_j(x)$ changing with large $x$ ($j=1,\ldots,m$) through modelling dependence within the asymptotic independence class. This model gives estimates of $\theta(x,m)$ that can be constant or vary with $x$, $x>u$. This additional flexibility comes at a price: inference is required for up to $2m$ parameters, and for an arbitrary  $m$-dimensional distribution $G$.


The asymptotic arguments for the Heffernan--Tawn model are given in Section~\ref{sec:SUB}, for an $(m+1)$-dimensional variable with Laplace marginal distributions. Suppose that the monotone increasing transformation $T=(T_0,\ldots,T_m)$ transforms ${\bf X} = (X_0,\ldots,X_m)$ to ${\bf Y}=(Y_0,\ldots,Y_m)$, with  $Y_i=T_i(X_i)$  ($i=0,\ldots ,m$),  so that ${\bf Y}$ has Laplace marginal distributions.  In applications the Heffernan--Tawn model corresponds to a multivariate regression with
\begin{equation}\label{eq:multregression}
{\bf Y}_{-0}\mid \{Y_0=y\}
= \gras\alpha y + y^{\gras\beta}\gras Z
= \gras\alpha y + \gras\mu  y^{\gras\beta} + \gras\psi  y^{\gras\beta} {\bf Z^\ast}
\end{equation}
where here, and subsequently, the arithmetic is to be understood componentwise, with $\gras\alpha \in [-1,1]^m$, $\gras\beta \in [-\infty,1]^m$, 
$\gras\mu \in \mathbb R^m$,  $\gras\psi \in \mathbb R_+^m$,  and ${\bf Z}$ an $m$-dimensional random variable with ${\bf Z}\sim G$; ${\bf Z^\ast}$ has zero mean and unit variance for all marginal variables, with ${\bf Z^\ast}=(\gras Z-\gras\mu)/\gras\psi$. We require that \eqref{eq:multregression} holds for all $y>u$, where $u$ is a high threshold on the Laplace marginal scale. The parameters $(\gras\alpha, \gras\beta, \gras\mu, \gras\psi)$ determine the conditional mean and variance through
\[
\E({\bf Y}_{-0}\mid Y_0=y)= \gras\alpha y + \gras\mu  y^{\gras\beta},\quad
\var({\bf Y}_{-0}\mid Y_0=y)= \gras\psi^2  y^{2\gras\beta}.
\]
Thus $(\gras\alpha, \gras\beta, \gras\mu, \gras\psi)$ can be estimated by multivariate regression. The complication for inference is that the error distribution $G$ is in general unknown and arbitrary, apart from the first two moment properties mentioned above. One exception to this is when $\gras\alpha= \gras 1$ and $\gras\beta = \gras 0$, in which case  the Heffernan--Tawn model reduces to known asymptotically-dependent models with $G$ directly related to $H$, as detailed in Section~\ref{sec:SUB}.


\citet{HeffernanTawn} and \citet{EastoeTawn} used a stepwise inference procedure, estimating $(\gras\alpha, \gras\beta, \gras\mu, \gras\psi)$ under a working assumption that ${\bf Z^\ast}$ are independent normal variables. After obtaining parameter estimates, they estimated  $G$ nonparametrically using the empirical joint distribution of the observed standardised multivariate residuals, i.e., values of ${\bf Z}$ for $Y_0>u$.  There are weaknesses in this approach, which loses efficiency in the estimation of $(\gras\alpha, \gras\beta, \gras\mu, \gras\psi)$ and $G$ and in subsequent inferences due to the generally incorrect working assumption of normality. Moreover, as noted by \citet{PengQi}, the empirical estimation of $G$ leads to poor estimation of the upper tail of the conditional distribution of ${\bf Y}_{-0}\mid \{Y_0=y\}$, so it would be preferable to have a better, yet general, estimator of $G$.  Furthermore,  the uncertainty of the parameter estimation is unaccounted-for in the estimation of $G$ and of cluster functionals such as $\theta(x,m)$. 

\citet{ChengGillelandHeatonAghaKouchak} proposed a Bayesian approach to estimating the Heffernan--Tawn model in a single stage, but their estimation procedure involves changing the structure of the model and adding a noise term in \eqref{eq:multregression}, thereby allowing the likelihood term to be split appropriately. They also need strong prior information extracted from the stepwise inference method in order to get valid estimates for the model parameters, so this procedure does not really tackle the loss of efficiency of the stepwise estimation procedure.

We propose to overcome these weaknesses by using Bayesian semiparametric inference to estimate the model parameters and the distribution $G$, simultaneously performing the entire fitting procedure for the dependence model. This gives a new model for $G$, namely, a mixture of Gaussian distributions, which provides estimates of the conditional distribution of ${\bf Y}_{-0}\mid Y_0=y$ beyond the range of the current estimator and which in theory provides an arbitrary good approximation for $G$ \citep{MarronWand}. The Bayesian approach also provides a coherent framework for fitting a parsimonious parametric model; joint estimation of the model parameters enables the imposition of structure between them. For example, in multivariate problems the context may suggest that different components of $\gras\alpha$ may be identical. In the context of time series extremes, for first-order Markov models, it can 
be shown that $\gras\alpha$ and $\gras\beta$ involve at most two unknown parameters \citep{PapaStrokorbTawnButler}. Furthermore, when the ${\bf X}_{-0}$ are known to be asymptotically dependent on $X_0$, this method provides a new approach to modelling. 

We show the practical importance of the new approach by applying it to the daily mean flow time series  of the River Ray at Grendon Underwood in north-west London, with observations from October $1962$ to December $2008$. We consider only flows from October to March, as this period typically contains the largest flows and forms an approximately stationary series. An empirical estimate of $\theta(x,m)$, with $m=4$, is shown in Figure~\ref{fig:theta} with bootstrap-derived $95\%$ confidence intervals. A major weakness with this estimate is that it cannot be evaluated beyond the range of the data, so a model is needed to evaluate $\theta(x,m)$ for larger $x$. We select our modelling threshold $u$ to be the empirical $98\%$ marginal quantile of the data. Using the methods in \citet{EastoeTawn} we have an estimate of $\theta(x,4)$ for all $x>u$ using the stepwise estimation method. As seen in Figure~\ref{fig:theta}, this estimate converges to $1$ as $x\to x_F$ but we have no reliable method for deriving confidence intervals.  Figure~\ref{fig:theta} also shows posterior median estimates and $95\%$ credibility intervals obtained using our Bayesian semiparametric method.  These show broad agreement with both of the other estimates within the range of the data, but with tighter uncertainty intervals and statistically significant differences in extrapolation of $\theta(x,4)$, indicating that the new method has the potential to offer marked improvement for estimating $\theta(x,m)$ and other cluster functionals.

The paper is structured as follows.
We first briefly present the standard approaches to multivariate extremes in Section~\ref{sec:history}. We introduce the conditional model of interest in a multivariate framework in Section~\ref{sec:SUB}, followed by a section about modelling of dependent time series.  Section~\ref{sec:BAY} explains the Bayesian semiparametric inference procedure, which is used in Section~\ref{sec:simulation} to illustrate the efficiency gains of this new inference method on simulated data.
In Section~\ref{sec:data} we fit our model to the River Ray flow data and show its ability to estimate functionals of time series clusters other than the threshold-based index.

\section{Multivariate setup and classical models}\label{sec:history}


Both multivariate and time series extremes involve estimating the probability of events that may never yet have been observed. Suppose that ${\bf X}=(X_0,\ldots,X_m)$ is  an $(m+1)$-dimensional variable with joint distribution function $F_{{\bf X}}$ and  marginal distribution functions $F_0,\ldots,F_m$. We need to estimate the probability $\Pr({\bf X} \in A)$, where $A\subset\mathbb R^{m+1}$ is an extreme set, i.e., a set such that for all ${\bf x} \in A$, at least one component of ${\bf x}=(x_0, \ldots ,x_m)$ is extreme. To do this we must model $F_{\bf X}({\bf x})$ for all ${\bf x}\in B$, where $B$ is an extreme set that contains $A$. Let $A_i$ be the subset of $A$ for which component $i$ is largest on a quantile scale, i.e.,
\[
A_i= A\cap \big\{{\bf x} \in\mathbb R^{m+1} : F_{i} (x_i) >F_{j}(x_j),  j\in \{0, \ldots  ,  m\}\backslash \{i\}\big\},\quad i=0,\ldots ,m, 
\]
and let $v_i=\inf\{x_i: {\bf x} \in A_i\}$, so that we can write 
\begin{equation}\label{eq:prob_decomposition}
\pr({\bf X} \in A)= \sum_{i=0}^m \pr({\bf X}\in A_i\mid X_i>v_i)\pr(X_i>v_i).
\end{equation}
Thus estimates of marginal and dependence features are required for estimating the conditional probabilities in the sum and marginal distributions determine the second terms of the products in the sum. 

Although our approach applies to any form of set $A$, to focus our arguments we restrict ourselves to identical marginal distributions $F$ with upper endpoint $x_F$ and we set 
\[
A = A_0=\{X_0>x, X_1\leq x, \ldots ,X_m\leq x\},
\]
so $v_0=x$ and we estimate only the conditional probability term in \eqref{eq:prob_decomposition}, i.e., $\theta(x,m)$, as defined in \eqref{theta-x-m.eq}.


Early approaches to modelling the conditional distribution appearing in \eqref{theta-x-m.eq} assumed that $\gras X$ lies in the domain of attraction of a multivariate extreme value distribution \citep{ColesTawn94,deHaan} and applied these asymptotic models above a high threshold. Unlike in the univariate case, there is no finite parametrisation of the dependence structure; it can only be restricted to functions of a distribution $H$ on the $m$ unit simplex $S_m$ with $\int_{S_m} w_idH({\bf w})=(m+1)^{-1}$ ($i=0,\ldots ,m$),  where ${\bf w}=(w_0, \ldots ,w_m)$.  Both parametric and non-parametric inference for this class of models has been proposed. Numerous parametric models are available \citep[\citeauthor{Kotz.Nadarajah:2000}, \citeyear{Kotz.Nadarajah:2000}, Ch.~3;][]{CooleyDavisNaveau,BallaniSchlather}. Nonparametric estimation is also widely studied, mostly based on empirical estimators \citep{deHaan, HallTajvidi, EinmahldeHaanPiterbarg, EinmahlSegers}.

A major weakness of these early methods is that for these models either $\chi_j>0$ or $(X_0,X_j)$ is independent for all $j=1,\ldots,m$, whereas there are distributions, such as the multivariate Gaussian, with $\chi_j=0$ but $(X_0,X_j)$ dependent. If $\chi_j>0$ for any $j=1,\ldots,m$, these models give estimates of $\theta(x,m)\to c_m$ as $x\to x_F$, where $c_m<1$. This class of models is not flexible enough to cover distributions that are dependent at finite levels, but asymptotically independent for all pairs of variables.


\section{Threshold-based model for conditional probabilities}\label{sec:SUB}

\subsection{Heffernan--Tawn model}
In order to provide a model characterising conditional probabilities, such as those which describe the clustering behaviour, we need a model for multivariate extreme values, and in particular we require a model for the joint distribution of $X_1,\ldots,X_m$ given that $X_0>x$. In this section we suppose that the marginal distributions of $\gras X$ are not identical, and that $X_0\sim F_0$ is in the domain of attraction of a generalised Pareto distribution, i.e., there exists a function $\sigma_u>0$ such that as $u\to x_{F_0}$, $(X_0-u)/\sigma_u$, conditional on $ X_0>u$, converges  to a generalised Pareto variable with unit scale parameter and shape parameter $\xi$.

The joint distribution is modelled via the marginal distributions $F_j$ and a copula for the dependence structure. To study the conditional behaviour of extremes, the copula formulation is most transparent when expressed with Laplace marginals. Let $T_j$ denote the transformation of the marginal distribution of $X_j$ to the Laplace scale, i.e., 
\[
 T_j(X_j)=\begin{cases}
           \log\left\{2F_j\left(X_j\right)\right\},& X_j<F_j^{-1}(1/2),\\
           -\log\left[2\left\{1-F_j\left(X_j\right)\right\}\right],& X_j>F_j^{-1}(1/2); 
          \end{cases}\quad j=0,\ldots,m, 
\]
the specification of the $F_j$ is discussed in Section~\ref{sec:classical_inference}. Assume there exist $m$-dimensional functions $\gras a(x)=\{a_1(x),\ldots,a_m(x)\}$ and $\gras b(x)=\{b_1(x),\ldots,b_m(x)\}>{\gras 0}$ for which
\begin{equation}\label{eq:HTcondition}
\pr\left[\dfrac{T_j\left(X_j\right)-a_j\left\{T_0\left(X_0\right)\right\}}{b_j\left\{T_0\left(X_0\right)\right\}}\leq z_j,\, j=1,\ldots,m\, \middle\vert \, X_0>u\right]\rightarrow G(\gras z),\quad u\rightarrow x_{F_0},
\end{equation}
where all marginal distributions of $G$ are non-degenerate and $\gras z=(z_1,\ldots,z_m)\in\mathbb R^m$. Hereafter we write the standardised $X_j$, or residual, as
\begin{equation*}\label{eq:HTzed}
	Z_j = \dfrac{T_j\left(X_j\right) - a_j\left\{T_0\left(X_0\right)\right\}}{b_j\left\{T_0\left(X_0\right)\right\}},\quad X_0>u,\ j=1,\ldots,m.
\end{equation*}
Under assumption \eqref{eq:HTcondition}, the rescaled conditioning variable $\left(X_0-u\right)/\sigma_u$ is asymptotically conditionally independent of the residual $\gras Z=(Z_1,\ldots,Z_m)$ given $X_0>u$, as $u\to x_{F_0}$.  That is,
\begin{align}\label{eq:HTindependence}
\Pr\left\{\gras Z\leq\gras z, \right.&\left.\left(X_0-u\right)/\sigma_u> x\ \middle\vert\  X_0>u\right\}\nonumber\\
		&=\Pr\left\{\gras Z\leq\gras z\mid \left(X_0-u\right)/\sigma_u>x\right\}\pr\left\{\left(X_0-u\right)/\sigma_u>x\mid X_0>u\right\}\nonumber\\
		&\to G(\gras z)\, \bar K(x),\quad u\rightarrow x_{F_0},
\end{align}
where $\bar K$ is the generalised Pareto distribution survivor function (\ref{eq:GPDmodel}) with scale and shape parameters $(1,\xi)$ and $G$ is the limit distribution of the residuals.

Equation \eqref{eq:HTcondition} can be illustrated through the particular case when $\gras X$ is a centred multivariate Gaussian distribution with correlation matrix elements $\rho_{ij}$, $i,j=0,\ldots,m$, $i\neq j$. In this case we can derive $a_j(x)=\sign(\rho_{0j})\rho^2_{0j}x$ and $b_j(x)=x^{1/2}$, and $G(\gras z)$ is a centred multivariate Gaussian distribution with variances $\rho_{0j}^2(1-\rho_{0j}^2)$ and correlation matrix elements
\[
 \rho'_{ij} = \dfrac{\rho_{ij}-\rho_{0i}\rho_{0j}}{\sqrt{\left(1-\rho^2_{0i}\right)\left(1-\rho^2_{0j}\right)}},\quad i\neq j.
\]

\citet{HeffernanTawn} and \citet{KeefPapaTawn} showed that under broad conditions, the component functions of ${\gras a}(x)$ and ${\gras b}(x)$ can be modelled by 
\begin{equation*}
a_{j}(x) = \alpha_{j}x,\quad b_{j}(x) = x^{\beta_{j}},\quad -1\leq\alpha_j\leq 1,\quad -\infty<\beta_j\leq 1,\quad j=1,\ldots,m.
\end{equation*}
In terms of the dependence structure, $\alpha_{j}$ and $\beta_{j}$ reflect the flexibility of the model. It turns out that $(X_0,X_j)$ are asymptotically dependent only if $\alpha_{j}=1$, $\beta_{j}=0$, and then 
\[
	\chi_j=\lim_{x\to x_{F_0}}\pr\left\{T_j(X_j)>x\mid T_0(X_0)>x\right\}=\int_0^\infty \bar G_j(-z)e^{-z}dz;
\]
with $\bar G_j$ the $j$th marginal survivor function of $G$; if $-1<\alpha_{j}<1$, then  $(X_0,X_j)$ are asymptotically independent, with positive extremal dependence if $\alpha_j>0$, negative extremal dependence if $\alpha_j<0$, and with extremal near-independence if $\alpha_j=0$ and $\beta_j=0$.

We set $\beta_j\geq 0$,  as when $\beta_{j}<0$ all the conditional quantiles for $X_j$ converge to the same value as $X_0$ increases, which is unlikely in most environmental contexts.
If the conditioning threshold $u$ is high enough that the conditional probability on the left of (\ref{eq:HTcondition}) is close to its limit, then the Heffernan--Tawn model can be stated as
\begin{equation}\label{eq:HTmodel}
T_j(X_j) = \alpha_j T_0(x) + \left\{T_0(x)\right\}^{\beta_j}Z_j,\quad X_0=x>u,\quad  j=1,\ldots,m,
\end{equation}
where $(Z_1,\ldots,Z_m)\sim G$ is independent of $X_0$, and $G$ can be any distribution with non-degenerate margins.

\subsection{Existing inference procedure}\label{sec:classical_inference}

We now outline the approach to inference suggested by \citet{HeffernanTawn}. Consider a vector $(X_0,\ldots,X_m)$ whose marginal distributions $F_0,\ldots,F_m$ each lie in the domain of attraction of a generalised Pareto distribution. We estimate them using the semiparametric estimator of \citet{ColesTawn94}, 
\begin{equation*}
	\widehat F_j(x) =
	\begin{cases}
	\widetilde{F}_j(x), & \quad x<u,\\
	1-\left\{1-\widetilde{F}_j(u)\right\}{\left(1+\hat\xi_j\dfrac{x-u}{\hat\sigma_{u,j}}\right)}_+^{-1/\hat \xi_j}, & \quad x\geq u,\\
	\end{cases}
\end{equation*}
where $\widetilde F_j$ is the empirical marginal distribution function of $X_j$.  Here $\hat\sigma_{u,j}$ and $\hat\xi_j$ are maximum likelihood estimates based on all exceedances of $u$, ignoring any dependence; their variances can be evaluated by a sandwich procedure \citep{FawcettWalshaw07} or by a block bootstrap. The margins $F_j$ are transformed to the Laplace scale through the transformation $\hat T_j\left(X_j\right)$.

Estimation of the probability of any extreme set of interest involves inference for model~\eqref{eq:HTmodel} with the estimators of parameters of the dependence model assumed independent of the parameter estimators of the marginal distribution. This assumption has been found not to be restrictive in other copula inference contexts \citep{Genest1995}. \citet{HeffernanTawn} proposed a stepwise inference procedure for estimating the extremal dependence structure, based on the working assumption that the residual variables $Z_1,\ldots, Z_m$ are independent and Gaussian with means $\mu_1,\ldots, \mu_m$ and variances $\psi^2_1,\ldots, \psi^2_m$. This assumption allows likelihood inference based on the assumed marginal densities,
\begin{equation*}
	T_j\left(X_j\right)\mid \left\{T_0(X_0)=x\right\}\sim\mathcal N\left(\alpha_jx+x^{\beta_j}\mu_j,\ x^{2\beta_j}\psi^2_j\right),\quad x>u,\ j=1,\ldots,m.
\end{equation*}
The first step of their procedure consists of a likelihood maximisation performed separately for each $j$, giving estimates of $\alpha_j$, $\beta_j$ and the nuisance parameters $\mu_j$ and $\psi_j$. Additional constraints, arising from results of \cite{KeefPapaTawn}, lead to the likelihood function being zero for certain combinations of  parameters $(\alpha_j, \beta_j)$. Thus the maximisation is over a subset of $[-1,1]\times [0,1]$ for these two parameters.  These constraints ensure that the conditional quantiles of $T_j(X_j)\mid T_0(x)$ are ordered in a decreasing sequence for all large $x$ under fitted models corresponding to positive asymptotic dependence, asymptotic independence and negative asymptotic dependence respectively. For details of these constraints see \cite{KeefPapaTawn}, who show that imposing these additional constraints improves inference of the conditional extremes model. Given model \eqref{eq:HTmodel} and the estimates $\hat\alpha_j$ and $\hat\beta_j$, the second step of the estimation procedure involves multivariate residuals $\gras Z$ for each data point, using the relation
\begin{equation*}
\hat Z_j = \dfrac{\hat T_j\left(X_j\right)-\hat\alpha_{j}\hat T_0\left(X_0\right)}{\left\{\hat T_0\left(X_0\right)\right\}^{\hat\beta_j}},\quad j=1,\ldots,m,\quad X_0>u,
\end{equation*}
and hence constructing the joint empirical distribution function $\hat G\left(\gras z\right)$.

An estimator for $\pr(A\mid X_0>x)$ for any extreme set $A$ is obtained as follows: sample $R$ independent replicates $X_0^{(1)},\ldots,X_0^{(R)}$ of $X_0$ conditional on $X_0>x$ from a generalised Pareto distribution with threshold $x$; independently sample $\gras Z^{(1)},\ldots,\gras Z^{(R)}$ from the joint empirical distribution function $\hat G$; compute
\[
 \gras X_{-0}^{(r)}=\hat {\gras T}^{-1}\left[\gras{\hat\alpha} \hat T_0(X_0^{(r)})+\left\{\hat T_0({X_0^{(r)}})\right\}^{\gras{\hat\beta}} \gras Z^{(r)}\right],
 \quad r=1,\ldots,R,
\]
where vector arithmetic is to be understood componentwise and $\hat{\gras T}^{-1}=(T_1^{-1},\ldots,T_m^{-1})$ is a componentwise back-transformation to the original scale; then the estimator for $\pr(A\mid X_0>x)$ is
\[
	\dfrac{1}{R}\sum_{r=1}^R I\left\{\left(X_0^{(r)},\gras X_{-0}^{(r)}\right)\in A\right\},
\]
where $I$ is the indicator function.

In the rest of the paper, we are interested in estimating the conditional probability $\theta(x,m)$, corresponding to $A=\{X_0>x, X_1<x,\ldots,X_m<x\}$, for which the Heffernan--Tawn model provides a characterisation. Under the assumption that the limit (\ref{eq:HTcondition}) approximately holds for some subasymptotic $u$, \citet{EastoeTawn} obtain 
\begin{equation}\label{eq:thetaIntegral}
	\theta(x,m)=\int_x^\infty G\{\gras z(x,y)\}k_x(y)\d y,\quad x>u,
\end{equation}
where $k_x(y)$ is the generalised Pareto density for threshold $x$, with scale parameter $1 +\xi (x-u)$ and shape parameter $\xi$, and $\gras z(x,y)$ is an $m$-dimensional vector with elements
\begin{equation}\label{eq:HTsmallzed}
	z_j\left(x,y\right) = \dfrac{T_j\left(x\right)-\alpha_j T_0\left(y\right)}{\left\{T_0\left(y\right)\right\}^{\beta_j}},\quad j=1,\ldots,m.
\end{equation}
A Monte Carlo approximation  to the integral~\eqref{eq:thetaIntegral} gives the estimator 
\[
	\hat\theta(x,m)=
	\dfrac{1}{R}\sum_{r=1}^R\hat G\left\{\gras z\left(x,X_0^{(r)}\right)\right\},
\]
where $\gras z(x,X_0^{(r)})$ is given by expression (\ref{eq:HTsmallzed}) with $\gras\alpha$ and $\gras\beta$ replaced by estimates. Monte Carlo variability can be reduced by using the same pseudo-random sequence when generating samples for different values of $x$. \citet{EastoeTawn} use a bootstrap method to get confidence bounds for $\hat\theta(x,m)$, but \citet{Carvalho} found this to be unreliable.

Four main weaknesses of this inference procedure justify developing a more comprehensive approach.  First, the working assumption needed for the likelihood maximisation is that the residuals have independent Gaussian distributions, and it is hard to quantify how this affects inference. Second, ignoring the variability of $\hat{\gras\alpha}$, $\hat{\gras\beta}$ estimated in the first step leads to underestimation of the uncertainty in the estimate for the residual distribution $G$ and hence also for $\theta(x,m)$. Third, the empirical estimation of $G$ restricts estimates of extremal conditional probabilities, as simulated $\gras Z$ values provide no extrapolation over observed values of $\gras Z$. Fourth, the inability to impose natural constraints on $\left(\alpha_1,\ldots,\alpha_m\right)$ and $\left(\beta_1,\ldots,\beta_m\right)$ leads to inefficiency.

\section{Modelling dependence in time}\label{sec:ts}
Consider a stationary time series $\{X_t\}$ satisfying appropriate long-range dependence properties and  with marginal distribution $F$. The threshold-based extremal index $\theta(x,m)$ summarises the key extremal dependence in time series. In the block-maxima context, the distribution of the block maximum $M_n=\max\{X_1,\ldots,X_n\}$ at a level $x$ is approximately $\{F(x)\}^{n\theta(x,m)}$ for large $x$, $n$ and $m$ \citep{OBrien,KratzRootzen97}. The associated independent series $\{X^\ast_t\}$, having the same marginal distribution as $\{X_t\}$ but independent observations, has $M_n^\ast=\max\{X_1^\ast,\ldots,X_n^\ast\}$ with distribution function $\{F(x)\}^n$. So $\pr(M_n<x)\approx\{\pr(M^\ast_n<x)\}^{\theta(x,m)}$, with $\theta(x,m)$ accounting for the dependence.

The most popular approach to dealing with short-range dependent in such series is the peaks over threshold (POT) approach formalised by \citet{DavisonSmith}. This approach consists of selecting a high threshold $u$, identifying independent clusters of exceedances of $u$, picking the maximum $Y$ of each cluster, and then fitting to these cluster maxima the generalised Pareto distribution 
\begin{equation}\label{eq:GPDmodel}
 \pr(Y<x\mid Y>u) = 1-\left(1+\xi\dfrac{x-u}{\sigma_u}\right)_+^{-1/\xi},\quad x>u.
\end{equation}
The limiting results are used as an approximation for data at subasymptotic levels with limit distribution \eqref{eq:GPDmodel} taken as exact above a selected value of $u$.

Alternatives to the POT approach include modelling the series of all exceedances, for example using a Markov chain \citep{Smith.Tawn.Coles:1997, WinterTawn}, but they depend heavily on the validity of the underlying modelling assumptions and so may be inappropriate. 

\citet{EastoeTawn} consider the threshold-based extremal index as part of a model for the distribution of cluster maxima. Specifically they show that, for a given high threshold $u$, the cluster maxima, defined by the runs method with run-length $m$, have approximate distribution function
\begin{equation}\label{eq:eastawn_distribution}
 1-\dfrac{\theta(x,m)}{\theta(u,m)}\left(1+\xi\dfrac{x-u}{\sigma_u}\right)_+^{-1/\xi},\quad x>u,
\end{equation}
where the parameters $\xi$ and $\sigma_u>0$ determine the marginal distribution of the original series, and $x_+=\max(x,0)$. \citet{EastoeTawn} show how using the information in $\theta(x,m)$ can improve over the POT approach.
Distribution \eqref{eq:eastawn_distribution} reduces to the generalised Pareto model asymptotically as $u\to x_F$, and more generally when $\theta(x,m)=\theta(u,m)$ for all $x>u$. When estimates of $\theta(x,m)$ vary appreciably above $u$, this equality condition for $\theta(x,m)$ provides a diagnostic for situations where the POT method is inappropriate.

In our approach, when a Markov property can reasonably be assumed for a time series, the $\alpha_j$ and $\beta_j$ have a structure that we want to exploit.
\citet{PapaStrokorbTawnButler} and \citet{KulikSoulier} characterise the form of $a_j(x)$ and $b_j(x)$ under very weak assumptions. If the conditions needed for the Heffernan--Tawn simplification --- $a_1(x)=\alpha_1 x$ and $b_1(x)=x^{\beta_1}$ --- hold, then for positively associated first order Markov processes, \citet{PapaStrokorbTawnButler} show that either $(\alpha_j,\beta_j)=(1,0)$, or $(0,\beta^j)$ or $(\alpha^j,\beta)$ for some $\alpha\in [0,1)$ and $\beta=[0,1)$.  The first case corresponds to asymptotic dependence at all time lags, and the other two to different forms of decaying dependence under asymptotic independence. If $\{X_t\}$ follows an asymptotically dependent Markov process, then no parameters need be estimated, rather than $2m$. If the process is well-approximated by an asymptotically independent Markov process then either of the last two cases applies, and the number of parameters in the parametric component of the model reduces from $2m$ to $1$ or $2$.
In the case of a Gaussian AR(1) process $(X_t)$ with standard Gaussian margins
\[
 X_{t+1} = \rho X_t + \varepsilon_t,\quad
	  \varepsilon_t\stackrel{\rm iid}{\sim}\mathcal N\left(0,1-\rho^2\right),\ 
	  \rho\in (-1,1),
\]
\citet{HeffernanTawn} get normalising parameters $\alpha_j=\sign(\rho)\rho^{2j}$ and $\beta_j=1/2$, $j=1,\ldots,m$; the distribution $G(\gras z)$ is a centred multivariate Gaussian with variances $\rho^{2j}(1-\rho^{2j})$ and correlation matrix elements
\[
 \rho'_{ij} = \dfrac{\sign(\rho^{i+j})\rho^{j-i}\sqrt{1-\rho^{2i}}}{\sqrt{1-\rho^{2j}}},\quad i<j.
\]
See \citet{PapaStrokorbTawnButler} for many more examples of first order Markov processes and their resulting forms for $\alpha_j$, $\beta_j$ and $G$.


\section{Bayesian Semiparametrics}\label{sec:BAY}
\subsection{Overview}
Since the $m$-dimensional residual distribution $G$ in the Heffernan--Tawn model \eqref{eq:HTmodel} is unknown, the approach described in Section~\ref{sec:classical_inference} uses the joint empirical distribution function, which cannot model the tails of the conditional distribution of $X_j$ in \eqref{eq:HTmodel}.  Our proposed Bayesian approach instead takes $G$ to be a mixture of a potentially infinite number of multivariate Gaussian distributions through the use of a Dirichlet process. This approach can model any $G$ and capture its tails, and has the major benefit of allowing joint estimation of $\gras\alpha$, $\gras\beta$ and $G$.

Below we introduce Dirichlet processes and describe an approach to approximate Monte Carlo sampling from them. We then describe Bayesian semiparametric inference and the specification of prior distributions, and discuss implementation issues. Throughout we assume that we have $n$ observations from the distribution of $(X_1,\ldots,X_m)\mid X_0>u$, or equivalently from $\gras Z=(Z_1,\ldots,Z_m)$ if $\gras\alpha$ and $\gras\beta$ were known.

\subsection{Dirichlet process mixtures for the residual distribution}
Consider a bivariate problem, $m=1$; with known $\alpha$ and $\beta$. If we are to estimate the distribution function of $X_1\mid X_0>u$ this is equivalent to estimating the distribution function $G$ of the univariate random variable $Z_1\sim G$. If $G$ is estimated nonparametrically, its prior must be a distribution over a space of distributions. In this context, a widely used prior is the Dirichlet process \citep{Ferguson73} mixture. A simple model structure takes $G$ to be a mixture of an unknown number of distributions $Q_k$ having parameters ${\gras\lambda}_k$, $k=1,2,\ldots$, so that the Dirichlet process boils down to a distribution on the space of mixture distributions $P$ for $\{{\gras\lambda}_k\}$.
If ${\gras\lambda}_k\mid P\sim P$, then the distribution of $P$ is the Dirichlet process ${\rm DP}\left(\gamma P_0\right)$, where $P_0$ is the centre distribution and $\gamma>0$ the concentration parameter \citep{BayNonparametric}.

The definition of a Dirichlet process states that for any $p=1,2,\ldots$, and any finite measurable partition $\{B_1,\ldots,B_p\}$ of the space of the ${\gras\lambda}_k$,
\begin{equation*}
\left\{P(B_1),\ldots,P(B_p)\right\}\sim{\rm Dirichlet}\{\gamma P_0(B_1),\ldots,\gamma P_0(B_p)\}.
\end{equation*}
The interpretation of the Dirichlet process parameters stems from the properties
\begin{equation*}
 \E\{P(B_i)\} = P_0(B_i),\quad\var\{P(B_i)\} = \dfrac{P_0(B_i)\{1-P_0(B_i)\}}{\gamma+1},\quad i=1,\ldots,p,
\end{equation*}
so the ${\rm DP}(\gamma P_0)$ prior is closer to its mean $P_0$ and less variable for large values of $\gamma$. A constructive characterisation of the Dirichlet process is the stick-breaking representation \citep{Sethuraman}
\begin{equation}\label{eq:Sethuraman}
 P(\cdot) = \sum_{k=1}^\infty w_k\delta_{{\gras\lambda}_k}(\cdot),
\end{equation}
where $\delta_{{\gras\lambda}}$ denotes a distribution concentrated on ${\gras\lambda}$, and ${{\gras\lambda}}_1, {\gras\lambda}_2, \ldots$ are independent, $P_0$-distributed, and independent of the random weights $w_k\geq 0$, which satisfy $\sum_{k=1}^\infty w_k = 1$. The stick-breaking process takes its name from the computation of the weights: define $V_1,V_2,\ldots$ as the breaks independently sampled from a ${\rm Beta}(1,\gamma)$ distribution. The weights are then 
\begin{equation}\label{eq:stickbreaking}
 w_1 = V_1,\quad w_k = V_k\prod_{i=1}^{k-1}(1-V_i),\quad k=2,3,\ldots.
\end{equation}

\citet{IshwaranZarepour00} use formulation \eqref{eq:Sethuraman} to express the Dirichlet process in terms of the random variables $w_k$ and ${\gras\lambda}_k$. They also introduce index variables $c_1,\ldots,c_n$ that describe the components of the mixture in which the observations $z_1,\ldots,z_n$ lie, giving a stick-breaking representation in terms of the index variables $c_i$ rather than the random variables ${\gras\lambda}_k$.

A key step in deriving a posterior distribution is the truncation of the sum in the stick-breaking representation, i.e., replacing the infinite sum in \eqref{eq:Sethuraman} by a sum up to $N$. This is achieved by imposing $V_N=1$ in \eqref{eq:stickbreaking}.
The accuracy of the stick-breaking approximation improves exponentially fast in terms of the $\mathcal L_1$ error \citep{IshwaranJames01}, as
\begin{equation*}
{\Vert M_N-M_\infty\Vert}_1\leq 4\left[1-\E\left\{{\left(\sum_{k=1}^{N-1}w_k\right)}^n\right\}\right]\approx 4n\exp\left(-\dfrac{N-1}{\gamma}\right),
\end{equation*}
where $M_N$ is the marginal density 
\begin{equation*}
\int\prod_{i=1}^n\left\{\sum_{k=1}^N w_k Q_k\left(\d {Z}_i\mid{\gras\lambda}_i\right)\right\}{\rm DP}\left(\d P_N\right).
\end{equation*}
For example, the error is smaller than $10^{-29}$ when truncating the stick-breaking sum representation at $N=150$ as in our real data analysis (Section~\ref{sec:data}) for which $\gamma\leq 2$ and $n=154$.

Taking into account the transformations discussed above, a simple model for $G$ involving the Dirichlet process prior is
\begin{equation}\label{eq:DPtruncatedmodel}
\begin{split}
Z\mid c,\Lambda &\stackrel{{\rm ind}}{\sim} Q_c=Q(\cdot;{\gras\lambda}_{c}),\\
c\mid P_N &\stackrel{{\rm iid}}{\sim} P_N=\sum_{k=1}^N w_k\delta_k(\cdot),\\
\left(\Lambda, \gras w\right)   &\sim\pi_{\Lambda}(\cdot)\pi_{\gras w}(\cdot),
\end{split}
\end{equation}
where $\Lambda$ is the matrix with rows ${\gras\lambda}_1,\ldots,{\gras\lambda}_N$, and $\gras w=(w_1,\ldots,w_N)$, with $\sum_{k=1}^N w_k = 1$ for some suitably large $N$. To lighten the notation we write $Z_1\mid{\gras\lambda}_c$ instead of $Z_1\mid c,\Lambda$ in what follows. Taking the $Q_k$ ($k=1,\ldots,N$) to be normal distributions with means $\mu_{1,k}$ and variances $\psi^2_{1,k}$ leads to $\Lambda$ being a $2\times N$ matrix, with rows ${\gras\lambda}_k=(\mu_{1,k},\psi^2_{1,k})$. Model \eqref{eq:DPtruncatedmodel} is made more flexible by adding a hyperprior for the concentration parameter $\gamma$.

\subsection{Multivariate semiparametric setting}\label{sec:multivariate_semipar}
We now specify the features of our algorithm, finally yielding model~\eqref{eq:DPsemiparametricmodel}.
We must add a further element to \eqref{eq:DPtruncatedmodel}: covariates, i.e., the parametric part of the Heffernan--Tawn model \eqref{eq:HTmodel}, to recognise that $\gras\alpha$ and $\gras\beta$ are unknown. This is achieved using a covariate-dependent Dirichlet process, and it can be formulated in terms of the truncated stick-breaking representation as
\begin{equation}\label{eq:truncatedStickBreaking}
P_{\mid x}(\cdot) = \sum_{k=1}^N w_k\delta_{{{\gras\lambda}}_k(x)}(\cdot),
\end{equation}
so that a single output of the stick-breaking procedure gives rise to a whole family of distributions indexed by $x$. Our data are the $n$ observations from $m$-dimensional variables $\gras X_{-0}$, given $X_0$ is large.
We assume that, conditional on $ T(X_0)>u$,  $T(\gras X_{-0})$ has a mixture of multivariate normal distributions, $\sum_{k=1}^\infty w_k\mathcal N_m$,
where the mean vector $M_k(x)$ and the covariance matrix $\Psi_k(x)$ of the $k$th normal component depend on the value $x$ of $T(X_0)$. For parsimony, the variance matrix $\Psi_k(x)$ is taken to be diagonal with diagonal elements $\{\Psi_{1,k}(x),\ldots,\Psi_{m,k}(x)\}$, as the mixture structure is considered flexible enough to capture the dependence between the elements of $\gras X_{-0}$.

As we use the truncated version of the stick-breaking representation~\eqref{eq:Sethuraman}, the conditional distribution for the weights $w_i$ is a generalised Dirichlet distribution \citep{ConnorMosimann}, written as $\rm GDirichlet$.
This gives the final form of our semiparametric model:

\begin{align}\label{eq:DPsemiparametricmodel}
T(X_j)\mid \{T(X_0)=x,\ \alpha_j,\ \beta_j,\ M_{j,c}(x),\ \Psi_{j,c}(x)\} &\stackrel{\rm ind}{\sim}\mathcal N\left\{M_{j,c}(x),\Psi_{j,c}(x)\right\},\ j=1,\ldots ,m,\ x>u,\nonumber\\
M_{j,c}(x) = \alpha_j x + \mu_{j,c}x^{\beta_j}, &\quad\Psi_{j,c}(x)=x^{2\beta_j}\psi^{2}_{j,c},\\
c\mid \gras w & \sim \sum_{k=1}^N w_k\delta_k(\cdot),\nonumber
\end{align}
where the prior for $(\gras\alpha, \gras\beta, \gras\mu_k, \gras\psi_k)$, with $\gras\mu_k=(\mu_{1,k},\ldots,\mu_{m,k})$ and $\gras\psi_k=(\psi_{1,k},\ldots,\psi_{m,k})$, takes the form
\begin{equation*}
\begin{split}
\gras w\mid\gamma &\sim\text{GDirichlet}\left(1,\gamma,\ldots,1,\gamma\right),\\
\gamma &\sim\text{Gamma}\left(\eta_1,\eta_2\right),\\
\alpha_j &\stackrel{\rm iid}{\sim}\mathcal U(0,1),\quad 
\beta_j \stackrel{\rm iid}{\sim}\mathcal U(0,1),\quad j=1,\ldots,m,\\
\mu_{j,k} &\stackrel{\rm ind}{\sim}\mathcal{N}\left(0,\psi^2_{(\mu)}\right),\quad \psi^{2}_{j,k} \stackrel{\rm iid}{\sim}\text{Inv-Gamma}\left(\nu_{1,j},\nu_{2,j}\right),\quad j=1,\ldots ,m,\ k=1, \ldots ,N,
\end{split}
\end{equation*}
with positive hyper-parameters $(\eta_1,\eta_2, \psi_{(\mu)}^2, \nu_1,\nu_2)$. The \citet{KeefPapaTawn} conditions mentioned in Section~\ref{sec:classical_inference} are built into the likelihood terms for $\alpha_j$ and $\beta_j$, so the Metropolis--Hastings scheme systematically rejects proposals outside the support of the posterior.

\subsection{Implementation issues}
The semiparametric multivariate Bayesian model \eqref{eq:DPsemiparametricmodel} has an added benefit of allowing us to structure the parametric component of the model. Assuming $\gras X$ to be a first order Markov process yields different structures discussed in Section~\ref{sec:ts}. For example, the different forms of decaying dependence in the class of asymptotic independence can be modelled by setting the priors as $\alpha\sim\mathcal U(0,1)$ and $\beta\sim\mathcal U(0,1)$, independently. The appropriate structure can be determined using standard diagnostics, and if adopted in the modelling will lead to substantially improved efficiency. Imposing continuous priors on $\alpha$ and $\beta$ induces a restriction to the class of asymptotically independent series, but both parameters can be arbitrary close to the boundaries of their support, ensuring that the behaviour of $\theta(x,m)$ and the extremal structure of dependence of the series are not affected at the high levels of interest. A reversible jump procedure \citep{Green} could be added to the current algorithm in order to enable $\alpha$ and $\beta$ to have prior masses on the support boundaries to ensure positive posterior probability of asymptotic dependence, see \citet{ColesPauli2002} for an example of this type of construction.

The shape and scale for the prior variances of the components $\psi_{j,k}^2$ are taken to be $\nu_1=\nu_2=2$ to make the model prefer numerous components with smaller variances to a few dispersed components. The posterior distribution for $\gamma$ depends on the logarithm of the last weight in the truncation (cf. Appendix~\ref{sec:AppendixPosteriors}) and can be numerically unstable, so a vague gamma prior truncated at small values is needed to ensure convergence.
Conjugacy of the prior densities allows analytical calculation of the posterior distributions for all parameters in model \eqref{eq:DPsemiparametricmodel} except $\gras\alpha$ and $\gras\beta$, for which a Metropolis--Hastings step is needed. We use a regional adaptive scheme in \citet{RobertsRosenthal} to avoid the choice of specific proposal variances. The posterior densities are mainly derived from \citet{IshwaranJames02}, and are given in Appendix~\ref{sec:AppendixPosteriors}.


As noticed by \citet{Porteus}, the Gibbs sampler used for model \eqref{eq:DPsemiparametricmodel} leads to a clustering bias, because the weights do not satisfy the weak ordering $\E(w_1)\geq\cdots\geq\E(w_{N-1})$.
\citet{PapaRoberts} suggested two different label switching moves to improve the mixing of the algorithm. Components cannot be simply swapped, as this would change the joint distribution of the weights. Label switching is not to be understood in its exact sense within this framework: if a switch between two mixture components is proposed and accepted, then only their means and variances are swapped and the index variables $c_i$ of the data points belonging to these components are renumbered accordingly.  We use this approach and adapt it to our semiparametric framework.

The results presented in Sections~\ref{sec:simulation} and \ref{sec:data} are promising, but two aspects would benefit from improvement. Bayesian semiparametric inference provides a valuable approach to uncertainty in the Heffernan--Tawn model, but the procedure is not fully Bayesian, since the marginal distribution is fitted using  maximum likelihood estimation. With further work we could include the fit for the marginal distribution within the fit for the dependence structure, but we would have to account for the temporal dependence between observations in order not to introduce bias. The second possibility for improvement pertains to the sampling of $\alpha_j$ and $\beta_j$ in \eqref{eq:HTmodel}: the special cases corresponding to the boundaries of their support should correspond to Dirac masses, so reversible jump Markov chain Monte Carlo sampling \citep{Green} could be used.


\section{Simulation study}\label{sec:simulation}
\subsection{Bivariate data}\label{sec:bivariate}
We start by showing how the Bayesian semiparametric approach to inference can improve over the stepwise approach in a bivariate setting. The working assumption of Gaussianity for the residual variable $Z$ is key to the stepwise process, and if it fails badly then the stepwise approach may perform poorly
relative to the Bayesian semiparametric approach. To illustrate this we take $Z$ to have a bimodal density, either a mixture of Gaussian densities, or a mixture of Laplace densities. As the former is a special case of the structure of the mixture components in the dependent Dirichlet process, we may expect a clear improvement in that case, but it is less clear what to expect in the latter case. 

\begin{figure}[!t]
	\centering
	\includegraphics[clip=true, trim=0cm 0cm 0cm 1.5cm, height=6.5cm]{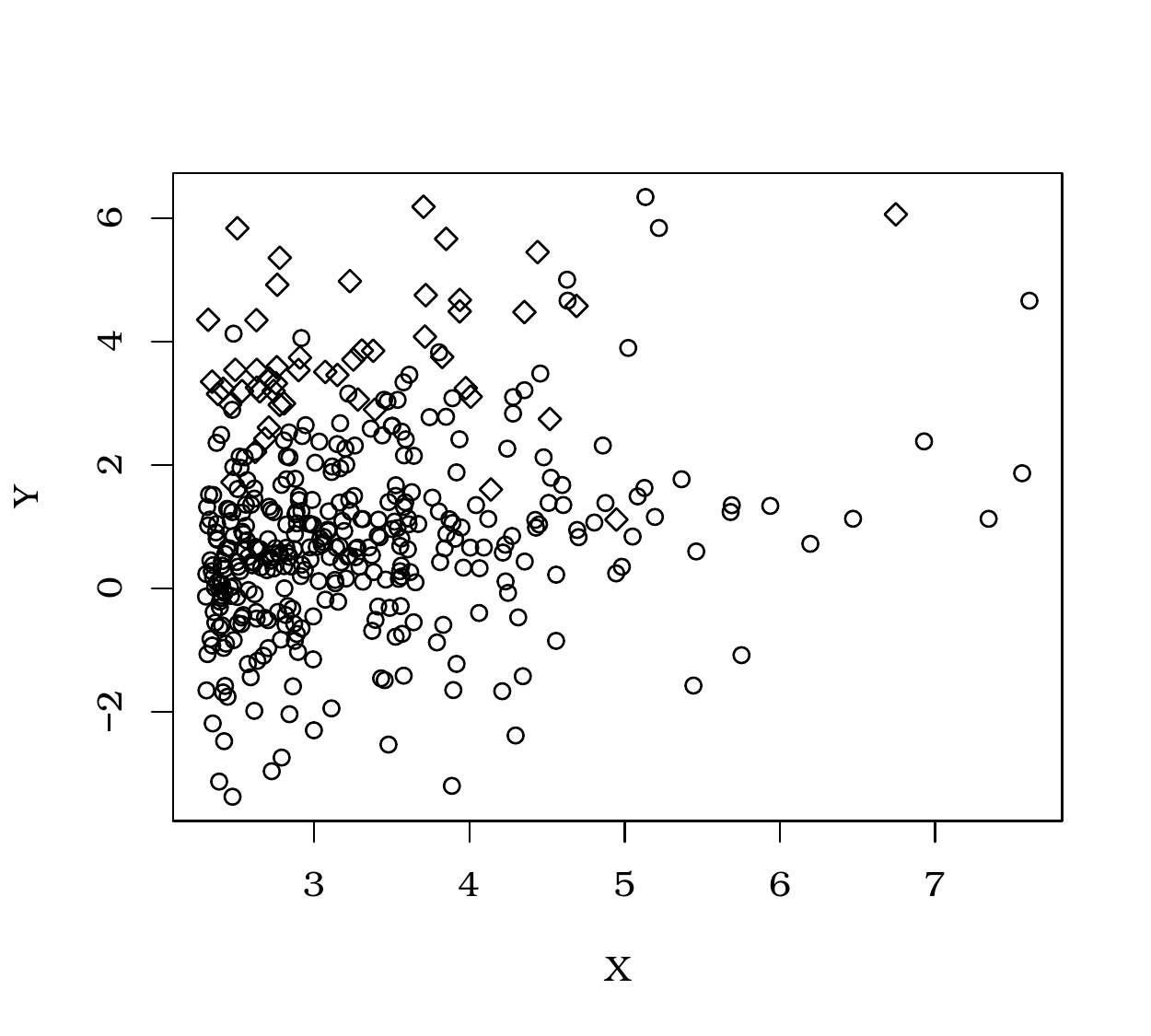}
	\includegraphics[clip=true, trim=0cm 0cm 0cm .9cm, height=6.5cm]{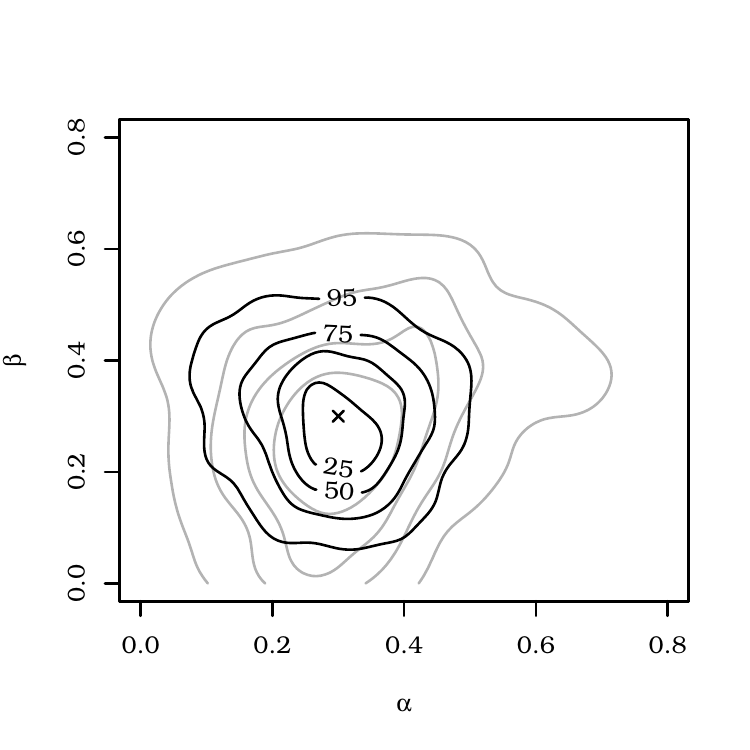}
	\caption{Results for simulation study based on the Heffernan--Tawn model with a bimodal residual distribution with two Laplace components.  Left panel: simulated data, with the two components shown as squares and circles. Right panel: kernel density estimate based on $1000$ estimates of $(\alpha,\beta)$ from the stepwise method (grey) and from the Bayesian semiparametric method (black) using the posterior medians as the summary statistic. The same contour levels are used for both density estimates. The true value is shown as a cross.}\label{fig:simHT}
\end{figure}

We generated data $(X,Y)$ directly from the Heffernan--Tawn model with parameters $(\alpha, \beta)$ subject to $X>u$, for large $u>0$, as follows:
\begin{enumerate}
\item Simulate $X$ as $u+\mbox{Exp}(1)$;
\item Independently simulate $Z$ from the required mixture model;
\item Let $Y=\alpha X + X^\beta Z$.
\end{enumerate}
We selected the mixture for the bimodal distribution of $Z$ such that the simulated $(X,Y)$ data are split into two clear clusters for large $X$ (see left panel of Figure~\ref{fig:simHT} for an example). 

We simulated $1000$ data sets each with $400$ points, roughly twice the number of exceedances available in our river flow application, and fitted the conditional model using the stepwise and the Bayesian semiparametric approaches. We compare the methods through the relative efficiency, measured as the ratio of the root mean squared error (RMSE) for the Bayesian approach to the RMSE of the stepwise approach. The estimators we consider in order to compute the efficiency are the mode, the mean, and the median of the posterior distribution of $\alpha$ and $\beta$ for the Bayesian approach and the maximum likelihood estimators of $\alpha$ and $\beta$ for the stepwise approach.

The benefits of the Bayesian semiparametric approach are clearly found, with similar relative efficiencies whether $Z$ is simulated with a Gaussian or Laplace mixture. The relative efficiency is broadly $0.6$ for $\alpha$ and in the range $0.5-0.65$ for $\beta$ depending on which of the three summary measures of the posterior distribution is chosen. The posterior number of components in the mixture is concentrated around $2$ and $3$, so the Bayesian semiparametric approach seems to capture the distribution of $Z$ well.  Figure~\ref{fig:simHT} shows the joint sampling distribution  of the estimators of $(\alpha,\beta)$ based on the two inference methods. The contours are similar, but suggest that the Bayesian approach estimates the true $(\alpha,\beta)$ more precisely.

Of key importance is the practical implication of this improvement, which is more naturally measured in terms of improved performance for estimating the threshold-based extremal index. Specifically we estimate $\theta(x,1)=\pr(Y<x\mid X>x)$, which requires accurate estimation of the distribution of $Z$ as well as of $\alpha$ and $\beta$.  The relative efficiency for $\theta(x,1)$ is computed, where the true value for $\theta(x,1)$ is obtained from a huge simulation from the true model. The relative efficiency varies over $x$, with values of $0.95$ and $0.90$
for the $99\%$ and $99.9\%$ quantiles, suggesting slight improvements within the range of the data. The relative efficiency reduces to $0.69$ at the $99.99\%$ quantile, suggesting that the real benefits in the Bayesian semiparametric approach arise when we extrapolate.

\subsection{AR(1) process}
We now compare the performances of the empirical, the stepwise and the   Bayesian semiparametric  inference procedures in estimating the threshold-based extremal index of a stationary time series.  The data are generated from a first-order Markov process with Gaussian copula and exponential margins. This is equivalent to having a standard Gaussian AR(1) process and using the probability integral transform to obtain exponential marginal distributions. In Gaussian margins this process has lag $\tau$ autocorrelation $\rho_\tau=\rho^{\tau}$, where we consider the set of $\{-0.75,-0.5,\ldots,0.5,0.75\}$ for the true value of $\rho$.
For each of these values of $\rho$, the process is asymptotically independent, with extremal index $\theta=1$, but it exhibits dependence at any subasymptotic threshold when $\rho\neq 0$. The true value for $\theta(x,m)$ is evaluated by computing the ratio of multivariate normal integrals
\begin{equation}
\theta(x,m)= \pr(X_0>x,X_1<x,\ldots,X_m<x)/\pr(X_0>x).
\label{eqn:Ratio}
\end{equation}
using the methods of \citet{mvtnorm} and \citet{mvtnormPackage}. The use of exponential margins ensures that the GPD marginal model is exact for all thresholds, so any bias in the estimation of $\theta(x,m)$ can be attributed to inference for the dependence structure.  
A similar approach was taken by \citet{EastoeTawn}. 

The three methods are applied to $1000$ data sets of length $8000$, approximately the length of the winter flow data studied in Section~\ref{sec:data}. This procedure is repeated for each value of $\rho$ in the range $\{-0.75,\ldots,0.75\}$.
The empirical method simply estimates each of the probabilities in expression~\eqref{eqn:Ratio} empirically.  Often called the runs estimate \citep{SmithWeissman}, this is not defined beyond the largest value of the sample, whereas the other two methods do not suffer this weakness.
In each case the marginal threshold $u$ for the modelling and inference is fixed at the $95\%$ empirical quantile of each series.  Unlike the stepwise procedure, the Bayesian semiparametric approach enables us to constrain $\{(\alpha_j,\beta_j):j=1,\ldots ,m\}$, and this allows us to exploit our knowledge of the Markovian structure of the process to impose the 
constraints on the $\alpha_j$ and $\beta_j$ discussed in Section~\ref{sec:ts}, thus reducing the number of parameters from $2m$ to $1$ or $2$.

\begin{figure}[!t]
 \centering
 \includegraphics[clip=true, trim=0cm 0cm 0cm 1cm, width=10cm]{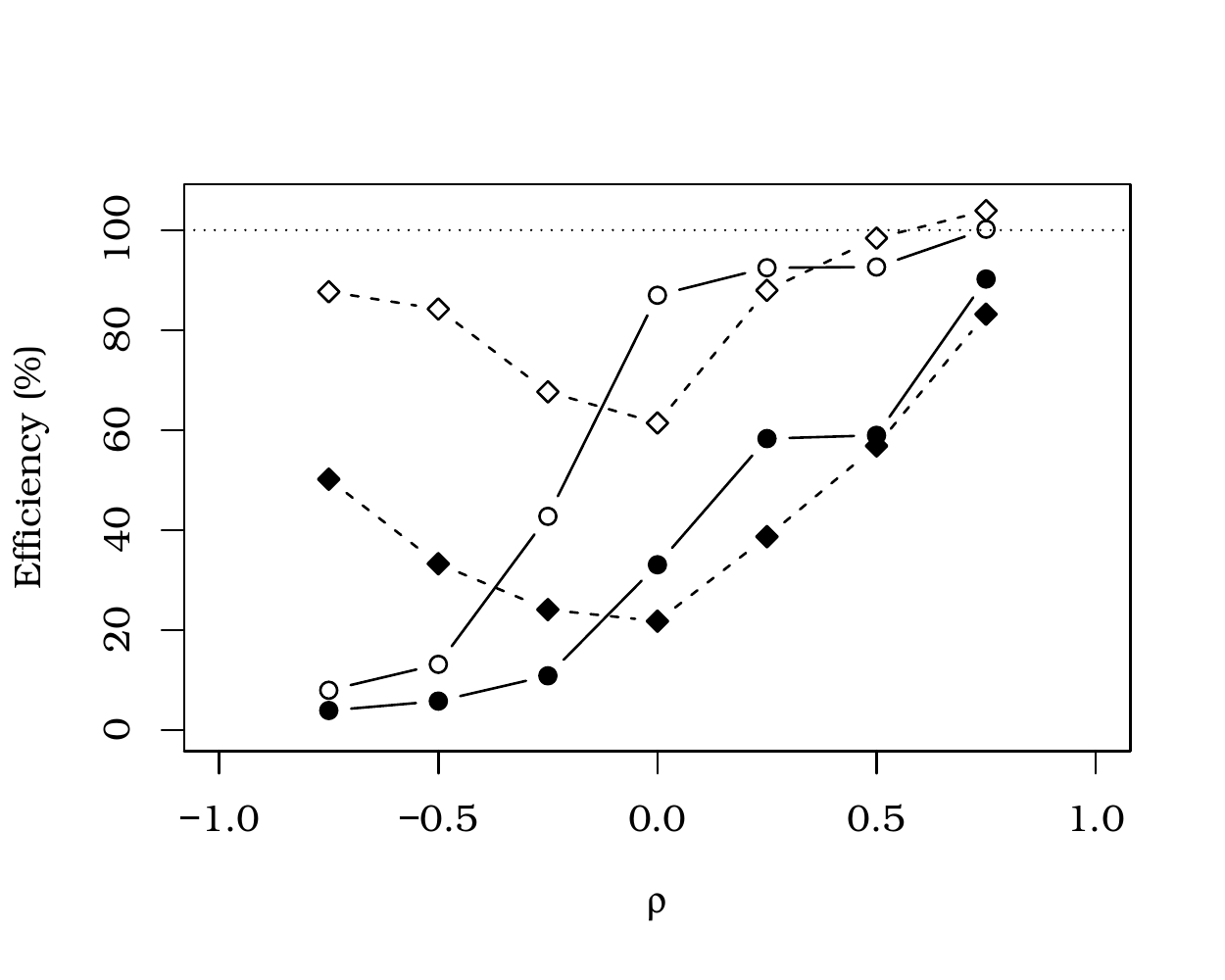}
 \caption{Ratios (\%) of RMSEs computed with estimates of $\theta(x,m)$, the numerator being the RMSE derived from the posterior median in our approach and the denominator being the RMSE corresponding to the stepwise estimate. The cases $m=1$ (circles) and $m=4$ (diamonds) are illustrated for $x$ at the $99\%$ (empty symbols) and $99.99\%$ (filled symbols) levels.}\label{fig:efficiencies}
\end{figure}

We estimate $\theta(x,m)$ for a range of high quantiles $x$ and for declustering run-lengths $m=1$ and $4$.
Table~\ref{tab:efficiencyAR1} shows the ratios of RMSEs of the posterior median of $\theta(x,m)$ from  the Bayesian semiparametric approach and the empirical and the stepwise estimators in the particular case when $\rho=0.5$.
The Bayesian semiparametric estimator is always superior to the empirical estimator, with the advantage improving as $x$ increases. For the stepwise approach the results are similar to those in Section~\ref{sec:bivariate}: the two estimators are similar at low levels but the Bayesian semiparametric estimator performs better at higher levels. Figure~\ref{fig:efficiencies} summarises the results for all values of $\rho$, showing a major improvement of our method over the stepwise approach for negative autocorrelation and short run-length, with increased gain at higher levels.
In order to assess the effectiveness of imposing the Markovian structure in the Bayesian semiparametric approach, we also fitted the $1000$ simulated time series with unconstrained $\gras\alpha$ and $\gras\beta$ in the case $\rho=0.5$. The efficiency of the unconstrained approach only declines relative to the constrained approach at high quantiles. For example the $68\%$  in the bottom right of Table~\ref{tab:efficiencyAR1} increases to $75\%$.

\begin{table}[!t]
\centering
\begin{tabular}{c|cc||cc}
        & \multicolumn{2}{c||}{m=1} & \multicolumn{2}{c}{m=4} \\
    Level & Empirical & Stepwise & Empirical & Stepwise \\
\hline
98\%     & 88 & 101 & 88 & 100 \\
99\%     & 68 &  92 & 71 &  98 \\
99.99\%  &  -- &  59 &  -- &  57 \\
\end{tabular}
\caption{Ratios (\%) of RMSEs computed with estimates of $\theta(x,m)$; the numerator of these efficiencies is always the RMSE estimate derived from the posterior median in the Bayesian semiparametric approach, and the denominator is either the RMSE corresponding to the runs estimate (Empirical) or to the stepwise estimate (Stepwise). Empty cells correspond to high levels of $x$ for which estimates of $\theta(x,m)$ cannot be evaluated.}\label{tab:efficiencyAR1}
\end{table}

\begin{figure}[!t]
 \centering
 \includegraphics[clip=true, trim=0cm 0cm 0cm 1cm, width=10cm]{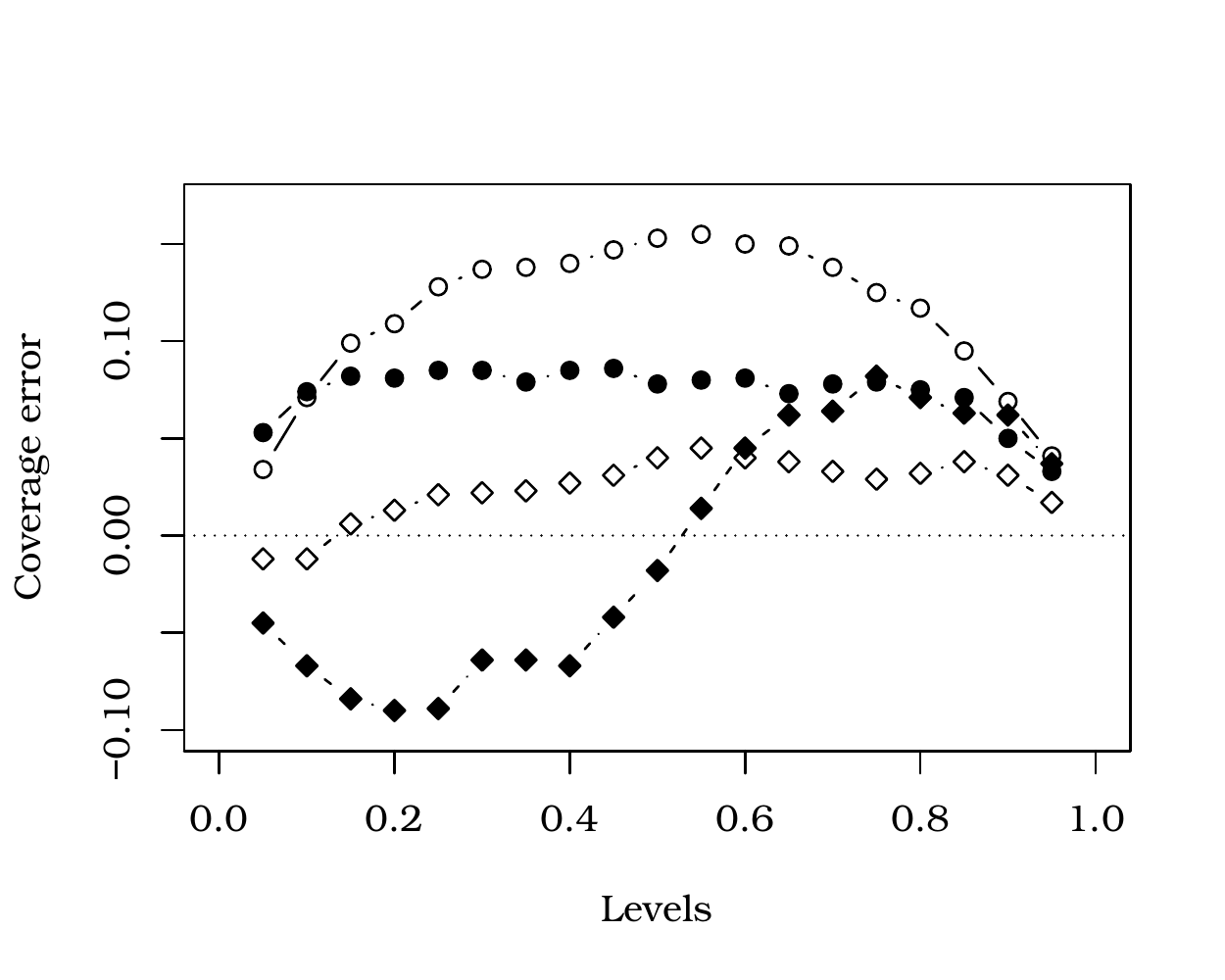}
 \caption{Coverage error for $\theta(x,m)$ computed for confidence levels $0.05, 0.1, \ldots, 0.95$, for $x=98\%$ (empty symbols) and $x=99.999\%$ (filled symbols), with $m=1$, for the stepwise approach (circles) and the Bayesian semiparametric approach (diamonds).}\label{fig:coverage}
\end{figure}

We expect the Bayesian approach to gain accuracy in terms of frequentist coverage of $\theta(x,m)$, as it fits the data in one stage and thus provides a better measure of uncertainty. To assess this we considered bootstrap confidence intervals for the stepwise method and credible intervals for the Bayesian method, both of the type $[L_{\alpha},\infty)$. Here $L_\alpha$ is the $\alpha$th quantile of the distribution of the estimator, considering bootstrap estimates for the former and posterior samples for the latter.
Using the same $1000$ simulated data sets as earlier in this section, we computed the proportion of times that the true value of $\theta(x,m)$ would fall in these confidence or credible intervals, for a range of $\alpha$-confidence levels from $5\%$ to $95\%$, different run-lengths $m$, and several levels $x$ for $\theta(x,m)$.
The coverage performance is summarised in Figure~\ref{fig:coverage}, which shows the difference between the calculated and the nominal coverage $\alpha$. Zero coverage error means perfect uncertainty assessment; positive and negative errors mean one-sided over- and under-coverage respectively. The stepwise approach over-estimates coverage for both levels of $x$ and all $\alpha$. At relatively low $x$-levels of $\theta(x,m)$, the gain in coverage accuracy by the Bayesian approach is remarkable in particular at mid-coverage levels, but it shows no marked improvement for larger $x$.

\section{Data analysis}\label{sec:data}
River flooding can badly damage properties and have huge insurance costs. Large-scale floods in the UK since the year $2000$ have caused insurance losses of \textsterling$5$ billion, and more than \textsterling$400$ million is spent each year on flood defences. Modelling the dependence of extreme water-levels is key to accurate prediction of flood risk.

Our application uses daily flows of the River Ray at Grendon Underwood, north-west of London, for the $47$ winters from $1962$ to $2008$. We assume stationarity of the series over the winter months.  River flows in this catchment are typically short-range dependent: after heavy rainfall the flow can reach high values before decreasing gradually as the river returns to its baseflow regime. We thus expect the flow to be dependent at extreme levels and at small lags, so a small run length $m$ is required. For illustration we take $m=1$, $7$, with the former being the more appropriate.

Standard graphical methods \citep{ismev2001} were used to choose the $95\%$ empirical quantile as the marginal threshold. A sensitivity analysis on a range of thresholds gave results  similar to those below. The Heffernan--Tawn model was then fitted to the data transformed to Laplace margins, with $u$ as the $98\%$ empirical quantile and $m=1$, 7. A higher threshold was selected for the dependence modelling to ensure the independence of $X_0$ and $\gras Z$ in approximation \eqref{eq:HTmodel}.

\begin{figure}[!t]
 \centering
 \includegraphics[clip=true, trim=0cm 0cm 0cm 1cm, width=9cm]{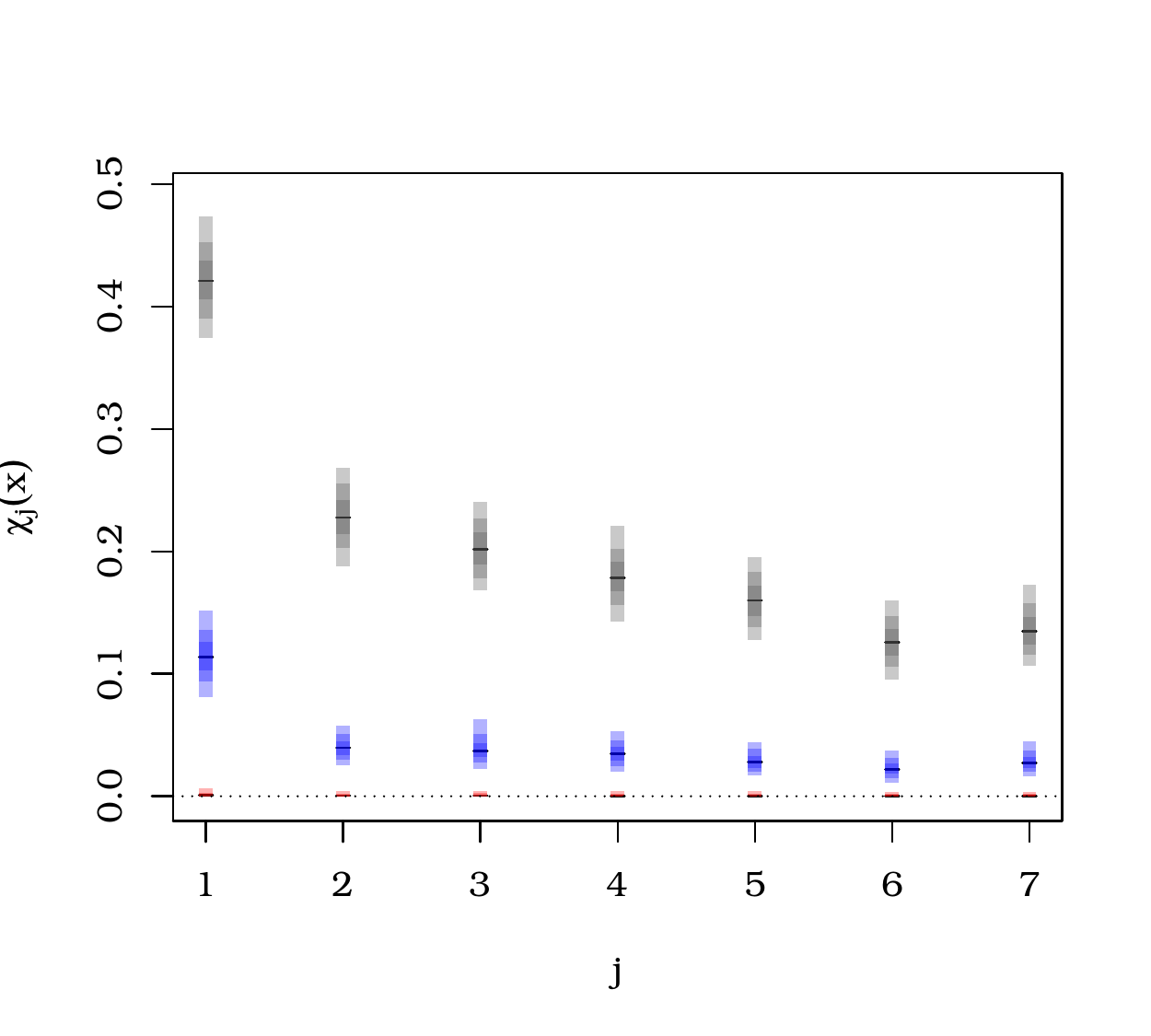}
 \caption{Posterior summaries for $\chi_j(x)$, $j=1,\ldots,7$, with $x$ the $95\%$ (grey), $99\%$ (blue), and $99.99\%$ (red) marginal quantiles of the River Ray data. The different shades of colour indicate the $95\%$, $80\%$, and $50\%$ highest density regions of the posterior densities. The black segments indicate the posterior medians.}\label{fig:chi}
\end{figure}

We first investigate the asymptotic structure of the data at different lags. We use $\chi_j(x)=\pr(X_j>x\mid X_0>x)$, $j=1,\ldots,m$, whose limit $\chi_j$ (cf.~Section~\ref{sec:intro}) either measures the degree of association within the asymptotic dependence class when $\chi_j>0$ or indicates asymptotic independence when $\chi_j=0$. A Monte Carlo integration similar to that used for estimating the posterior distribution of $\theta(x,m)$ is applied to get the posterior distribution of $\chi_j(x)$ for selected values of $x$ and $j=1,\ldots,7$ depicted in Figure~\ref{fig:chi}, where the posterior densities are summarised using highest density regions \citep{Hyndman}.
Convergence of $\chi_j(x)$ to $0$ at all lags is supported by the model. As expected we observe a monotone decay in extremal dependence over time lag. The flexibility of the conditional model is well illustrated here, as the procedure establishes positive dependence at any finite level but anticipates asymptotic independence of successive daily flows.

From the estimates of $\chi_j(x)$ we expect $\theta=1$ but $\theta(x,m)<1$ for $x<x_F$. We computed estimates of $\theta(x,m)$ for several values of $x$ based on the posterior distribution fitted with the Bayesian semiparametric approach and compared them with the stepwise approach and the empirical estimates.
We give block bootstrap confidence intervals for the stepwise and empirical estimates, with a block length ensuring that winters are not split between blocks. Figure~\ref{fig:theta}  shows estimates of $\theta(x,4)$ obtained with the three methods, with $m=4$ to show an intermediate estimate, as $\theta(x,7)\leq\theta(x,4)\leq\theta(x,1)$. The three methods broadly agree at historical levels, with wider confidence intervals for the runs estimate. For higher $x$, the stepwise estimate predicts stronger dependence than does the Bayesian estimate.

Table~\ref{tab:thetaData} shows that the three methods agree closely for $m=1$, with the posterior distribution giving slightly tighter credible intervals than the two other methods. For $m=7$,  the Bayesian approach seems to improve a little on the stepwise estimates when compared to the empirical estimates at low levels, partly because of  the Markov constraints on the $\alpha_j$ and $\beta_j$, which also reduce the uncertainty. In terms of convergence of $\theta(x,m)$ to the extremal index $\theta$, we observe that at very high levels and for both values of $m$, the estimates of $\theta(x,m)$ tend to the same values, which indicates coherence in the approach. This also illustrates the lesser concern of the choice of the run-length when we are interested in tail probabilities, typically when estimating cluster maxima distributions.

\begin{table}[tbp]
    \centering
    \[\begin{array}{l||ccc|ccc}
                    & \multicolumn{3}{c}{m=1}  & \multicolumn{3}{|c}{m=7}\\
        \text{Level} & \text{Empirical} & \text{Stepwise} & \text{Bayesian} & \text{Empirical} & \text{Stepwise} & \text{Bayesian}\\ \hline
    95\%     & 62_{(58,68)} & 62_{(61,70)} & 62_{(57,66)} & 35_{(33,41)} & 36_{(33,42)} & 33_{(27,38)}\\
    99\%     & 90_{(83,95)} & 87_{(83,91)} & 88_{(83,91)} & 80_{(70,88)} & 69_{(66,79)} & 76_{(71,80)}\\
    99.9\%   & \text{--}       & 96_{(93,99)} & 97_{(94,99)} & \text{--}       & 90_{(86,96)} & 98_{(96,99)}\\
    99.99\%  & \text{--}       & 99_{(96,100)}&100_{(98,100)}& \text{--}       & 98_{(93,100)}& 100_{(99,100)}\\
    \end{array}\]
    \caption{Estimation of the threshold-based extremal index $\theta(x,m)$ (\%) for four different levels of $x$ and $m=1,7$ on the Ray River winter flow data, with $95\%$ confidence intervals (CI) given as subscripts. Empirical: runs estimator (block bootstrap CI); Stepwise: Heffernan--Tawn method (block bootstrap CI); Bayesian: posterior median from the Bayesian semiparametric approach (quantiles of the posterior distribution).}\label{tab:thetaData}
\end{table}

The Bayesian semiparametric approach appears to offer a more coherent basis for the extrapolation to the required levels and uncertainty quantification for design purposes than does the stepwise method. Although we assessed the performance of Bayesian semiparametric estimates of $\theta(x,m)$, but other conditional probabilities could also be estimated using this  approach.



\clearpage
\appendix
\section{Posterior densities for the semiparametric model}\label{sec:AppendixPosteriors}
\paragraph{Posterior density for $\gras\mu_k$:} The posterior density for $\gras\mu_k=(\mu_{1,k},\ldots,\mu_{m,k})$ is multivariate Gaussian with independent margins, i.e., 
    \begin{equation*}
    \mu_{j,k}\mid \gras X_j,\gras X_0,\psi^2_{j,k},\alpha_j,\beta_j\stackrel{{\rm ind}}{\sim}\mathcal N\left(M_{(\mu_{j,k})}, S^2_{(\mu_{j,k})}\right),\quad j=1,\ldots,m,\quad k=1,\ldots,N,
    \end{equation*}
  with posterior mean and variance
    \begin{equation*}
    M_{(\mu_{j,k})}= S^2_{(\mu_{j,k})}\left(\dfrac{1}{\psi^2_{j,k}}\sum_{i\in C_k}\dfrac{X_{j,i}-\alpha_jX_{0,i}}{X_{0,i}^{\beta_j}}\right),\quad S^2_{(\mu_{j,k})}=\left(\dfrac{n_k}{\psi^2_{j,k}}+\dfrac{1}{\psi^2_{(\mu),j}}\right)^{-1},
    \end{equation*}
    where $\gras X_j=(X_{j,1},\ldots,X_{j,n})$ are the observations at the $j$th lag, $C_k=\{i:c_i=k\}$, and $n_k=\vert C_k\vert$ is the number of observations in component $k$; the $\psi^2_{(\mu),j}$ are the variance parameters of the prior for the components' means.
  \paragraph{Posterior density for $\gras\psi^2_k$:} The multivariate posterior density for the components' variances can be split into independent parts,
    \begin{equation*}
    \psi^2_{j,k}\mid\gras X_j,\gras X_0,\mu_{j,k},\alpha_j,\beta_j\stackrel{{\rm ind}}{\sim}\text{Inv-Gamma}\left(N_{1,j,k},N_{2,j,k}\right),\quad j=1,\ldots,m,\quad k=1,\ldots,N,
    \end{equation*}
  with parameters
    \begin{equation*}
    N_{1,j,k}= \dfrac{n_k}{2}+\nu_{1,j},\quad N_{2,j,k}=\dfrac{1}{2}\sum_{i\in C_k}\dfrac{\left(X_{j,i}-\alpha_jX_{0,i}-\mu_{j,k}X_{0,i}^{\beta_j}\right)^2}{X_{0,i}^{2\beta_j}}+\nu_{2,j}.
    \end{equation*}
  \paragraph{Posterior density for $\gras c$:} The posterior density is such that
    \begin{equation*}
    c_i\mid\gras X,\gras\mu,\gras\psi^2,\gras\alpha,\gras\beta,\gras w\stackrel{{\rm ind}}{\sim}\sum_{k=1}^N W_{k,i}\delta_k,\quad i=1,\ldots,n,
    \end{equation*}
  where here for convenience $\gras X$, $\gras\mu$ and $\gras\psi^2$ are the matrices with rows $(\gras X_0,\ldots,\gras X_m)$, $(\gras\mu_1,\ldots,\gras\mu_N)$, and $(\gras\psi^2_1,\ldots,\gras\psi^2_N)$ respectively; the stick-breaking weights are defined as
    \begin{equation*}
    W_{k,i}= \dfrac{w_k}{\bar W_i}\prod_{j=1}^m\left[\dfrac{1}{X_{0,i}^{\beta_j}\psi_{j,k}}\exp\left\{-\dfrac{1}{2}\dfrac{\left(X_{j,i}-\alpha_jX_{0,i}-\mu_{j,k}X_{0,i}^{\beta_j}\right)^2}{X_{0,i}^{2\beta_j}\psi^2_{j,k}}\right\}\right],
    \end{equation*}
  with $\bar W_i=\sum_{k=1}^N W_{k,i}$  ($i=1,\ldots,n$) constants that make the weights sum to $1$.
  \paragraph{Posterior density for $\gras w$:} The posterior density for $\gras w$ is generalised Dirichlet, 
    \[
		\gras w\mid\gras c,\ \gamma\sim {\rm GDirichlet}(a_1,b_1,\ldots,a_{N-1},b_{N-1}),
    \]
    where
    \[
		a_k= 1+n_k,\quad b_k= \gamma+\sum_{j=k+1}^Nn_j,\quad k=1,\ldots,N-1.
    \]

  \paragraph{Posterior density for $\gamma$:} The posterior density for the concentration parameter $\gamma$ is
    \begin{equation*}
	    {\rm Gamma}\left(N+\eta_1-1,\dfrac{\eta_2}{1-\eta_2\log w_N}\right)\gras 1_{[\varepsilon,\infty)},
    \end{equation*}
    with $\varepsilon>0$, typically $\varepsilon=0.5$, and $\gras 1$ is the indicator function.

\bibliographystyle{CUP}
\bibliography{bibliography}
\end{document}